\documentstyle[psfig]{caosp}
\begin{document}

\pubyear{1998}
\volume{27}
\firstpage{350}

\hauthor{J. Landstreet}

\title{Spectrum synthesis of sharp-lined A and B stars}

\author{J. D. Landstreet}

\institute{University of Western Ontario, London, Ontario, Canada, \\
and Observatoire Midi-Pyr\'{e}n\'{e}es, Toulouse, France}

\date{\today}
\maketitle

\begin{abstract}
I have carried out spectrum synthesis of $R = 120,000$ spectra of
several A and B stars having $v \sin i$ less than about 6 km
s$^{-1}$. The following conclusions emerge: (1) As $T_e$ descends from
12,000 to 8,000 K, microturbulent velocity $\xi$ deduced from
abundance analysis rises steadily from 0 to about 5 km s$^{-1}$. (2)
Stars with $\xi \geq 1$ km s$^{-1}$ show direct evidence in their line
profiles of the presence of macroscopic gas motions in the form of
line asymmetry (bisector curvature) which grows with increasing
$\xi$. (3) Above $T_e \approx 9000$ K, both weak and strong spectral
lines can be modelled with reasonable accuracy by conventional LTE
spectrum synthesis with a single assumed model atmosphere, abundance
table, $v \sin i$, and an appropriate (constant) value of $\xi$. (4)
In contrast, at $T_e \approx 8,000$ K the weak spectral lines are much
narrower than the strong lines. If the synthesis model is constrained
in $v \sin i$ and $\xi$ by the weak lines, {\em no} satisfactory model
can be found for the strong spectral lines. Consequently, chemical
abundances for such stars based only on strong lines may be
significantly in error.

\end{abstract} 

In cool stars, we learn a lot about conditions in the stellar
atmosphere from studies of the {\em shape} of line profiles, which
often contain much information about the velocity fields present. No
such studies exist for non-magnetic A and B stars because most of them
rotate so rapidly that the profiles contain very little information
except about the global rotation.

With current spectrographs and detectors, it is practical to search
for and observe A and B stars of $v \sin i \leq 6$ km s$^{-1}$, in
which information about local photospheric velocity fields is
detectable in the line profiles.  I have observed about a dozen such
stars with R = 120,000 with the Aur\'{e}lie spectrograph at OHP
or with the f/4 spectrograph at CFHT. The observed spectra are
modelled by spectrum synthesis.

This project has two main goals. The first is to test the adequacy of
the LTE models usually used to study A and B stars by seeing if a
single model (specified by $T_e$, $\log g$, rotational velocity $v
\sin i$, microturbulent velocity $\xi$, and abundance table) can
correctly reproduce the observed line profiles (which because of the
small values of $v \sin i$ contain much information about the local
stellar line profile) of both very weak and strong metal lines. The
second is to search for direct evidence in line profiles of convective
or other velocity fields, and to study the relationship of any such
evidence to the classical microturbulent velocity $\xi$ derived from
abundance analysis.

The observed spectra are modelled using a version of my spectrum
synthesis programme ZEEMAN. This programme calculates line profiles in
LTE, using a suitable Kurucz model atmosphere for each star. The
programme attempts to optimize the fit of the calculated profile to
the observed one, line by line, by adjusting $v \sin i$, radial
velocity $v_r$, $\xi$, and the abundance of the element producing the
line. All $gf$ values used have been carefully tested.

This paper presents results for two stars, HD
193452 (Sp = HgMn, $T_e$ = 10,500 K, $\log g$ = 4.0), and HD 108642 (Sp
= Am, $T_e$ = 8,100 K, $\log g$ = 4.1). 
For HD 193452, it is found both from Blackwell diagrams and from the line
profiles that $\xi \leq 1$ km s$^{-1}$ and $v \sin i \leq 2$ km
s$^{-1}$.  Comparison of the values of $\nabla_{rad}$ and
$\nabla_{ad}$ shows that only very mild convective instability is
expected, in agreement with the small $\xi$ found. The calculated line
profiles fit the observed lines extremely well, as may be seen in 
Figure 1 below showing representative fits to two iron lines.

For HD 108642, the {\em weak} lines show that both $v \sin i$ and
$\xi$ are $\leq 3$ km s$^{-1}$. However, Blackwell diagrams suggest a
value of $\xi$ near 4 km s$^{-1}$, and a comparison of the values of
$\nabla_{rad}$ and $\nabla_{ad}$ shows that the atmosphere should be
very unstable. {\bf Remarkably, the profiles of the strong lines are
much wider than the corresponding profiles of weak lines, and strongly
asymmetric}, with deeper blue wings than red; modelling of single
strong lines gives $v \sin i$ of about 6 -- 8 km
s$^{-1}$. Furthermore, calculation of line profiles for the strong
lines shows major discrepancies between the theoretical profiles
computed with constant $\xi$, and the observed profiles (see Figure
1). It is clear that the strong line profiles contain much information
about photospheric velocity fields which is not included in the simple
model used. It is also clear that abundances based on these strong
lines are likely not to be very accurate.

A number of other stars have also been modelled, with the following
results.  For stars with $T_e > 10,000$ K, $\xi$ is found to be below
about 1 km s$^{-1}$, and line profiles are very well described by LTE
synthesis even when $v \sin i \approx 1$ km s$^{-1}$. Below 10,000 K,
however, the deduced values of $\xi$ rise above 2 km s$^{-1}$, and the
line profiles are increasingly asymmetric, presumably due to
increasing velocities in the atmosphere. At 8000 K, the deduced $\xi
\approx 4$ km s$^{-1}$; the intrinsic line profiles of strong lines
are much wider, and more asymmetric, than those of weak lines, and
very poorly described by the spectrum synthesis. It seems clear that
classical analysis of these stars is a very rough first approximation
only. However, this profile structure is capable of furnishing much
further information about the nature of velocity fields present in
these A stars.

This work extends the studies of Gray (1992, chapter 18) and others on
line profile shapes of stars near the main sequence. Gray has shown
that main sequence stars show line asymmetry that changes with $T_e$;
cool stars have lines that are deeper on the red wing, while late A
stars have lines that show a deeper blue wing. My work includes stars
between about A5 and B9, and I find that the asymmetry in spectral
lines, when present, is always in the sense of a deeper blue line
wing, but that this asymmetry dies out at about 10,000 K. This
asymmetry suggests that convection in the photospheres of A
stars is composed mainly of small rising columns of hot gas balanced
by larger, slower downflows. This is opposite to the behaviour
observed in the sun, and to that predicted by numerical experiments
for A stars (e.g. Freytag et al 1996). It is not
clear what physical difference between convetion in A and G stars
could lead to such a fundamental difference in convective structure.
\vspace{+4mm}
\begin{figure}[ht]  
\psfig{figure=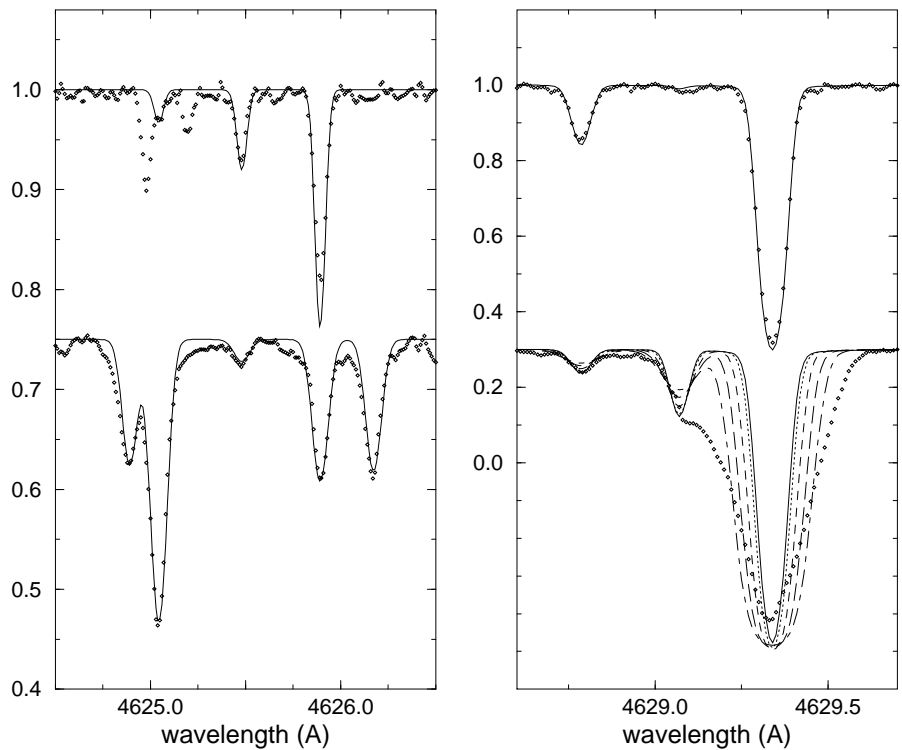,height=10.0cm}
\caption[]{Observed (triangles) and model (curves) line profiles
for two spectral regions of HD 193452 (upper profiles) and HD
108642 (lower profiles). For the Fe II line at 4629 \AA\ in HD 108642,
the model lines have $\xi$ increasing from 0 to 4 km s$^{-1}$. }
\end{figure}

\end{document}